\documentclass[12pt,aps,prl,showpacs]{revtex4}
\usepackage{graphicx}
\begin{document}

\bibliographystyle{aip}
\newcommand{\be}{\begin{equation}}
\newcommand{\ee}{\end{equation}}
\newcommand{\Be}{\begin{eqnarray}}
\newcommand{\Ee}{\end{eqnarray}}
\def\a{\alpha}
\def\b{\beta}
\def\g{\gamma}
\def\G{\Gamma}
\def\d{\delta}
\def\D{\Delta}
\def\e{\epsilon}
\def\k{\kappa}
\def\l{\lambda}
\def\L{\Lambda}
\def\t{\tau}
\def\om{\omega}
\def\Om{\Omega}
\def\s{\sigma}
\def\lg{\langle}
\def\rg{\rangle}\
\def\koff{k_{\rm off}}
\def\kon{k_{\rm on}}
\def\Keq{K_{\rm eq}}
\def\fext{f_{\rm ext}}

\title{Force-clamp spectroscopy of reversible bond breakage}
\author{Gregor Diezemann and Andreas Janshoff} 
\affiliation{Institut f\"ur Physikalische Chemie, Universit\"at Mainz,
Welderweg 11, 55099 Mainz, FRG}
\begin{abstract}
We consider reversible breaking of adhesion bonds or folding of proteins under the influence of a constant external force.  
We discuss the stochastic properties of the unbinding/rebinding events and analyze their mean number and their variance in the framework of simple two-state models. 
In the calculations, we exploit the analogy to single molecule fluorescence and particularly between unbinding/rebinding and photon emission events. 
Environmental fluctuation models are used to describe deviations from Markovian behavior.
The second moment of the event-number distribution is found to be very sensitive to possible exchange processes and can thus be used to identify temporal fluctuations of the transition rates.
\end{abstract}
\pacs{82.37.Np, 82.37.Rs, 87.10.Mn, 87.15.Fh}
\maketitle
\noindent
The unbinding of adhesion molecules or the unfolding of proteins can be monitored with dynamic force spectroscopy (DFS) on a single molecule level\cite{Evans:2001p105}. 
In many applications a linearly increasing load is applied to the system and the rupture force distribution is monitored as a function of the loading rate, yielding information about the details of the rupture event such as the possible existence of intermediates. 
Similar to other single molecule techniques, such as single molecule fluorescence (SMF) or single channel recording, DFS is able to disentangle complex reaction pathways\cite{Zhuang:2003p3322,Ritort:2006p3324}.
Two setups are usually employed to determine bond-strengths or bond-lifetimes.
In the linear loading mode, a force proportional to the observation time is applied, e.g. by moving the cantilever of an atomic force microscope away from the surface. 
Force-clamp spectroscopy (FCS), on the other hand, exerts a constant load on the bond, i.e. a fixed force is applied.
In the latter case, one observes individual unbinding/rebinding events of the molecular system as a function of time.
In both cases characteristic forces are in the pN-regime and bond-lifetimes are on the order of seconds\cite{Fernandez:2004p1274,Schlierf:2004p1358}. 
The theoretical models usually employed to describe the rupture process describe the dynamics in the direction of the reaction coordinate in terms of diffusive barrier 
crossing\cite{Schulten:1981p161}.
The rupture force or time is calculated from the mean first passage time in a potential consisting of the molecular potential and the harmonic potential describing the application of the force\cite{Bell:1978p2211,Hummer:2003p934,Dudko:2006p899}.
So far, predominantly irreversible rupture events have been investigated, both theoretically and experimentally.
More recently, reversible systems with finite rebinding rates came into 
focus\cite{Seifert:2002p163,Li:2006p164,Diezemann:2008subm}. 
Experimentally, reversible unfolding/refolding transitions have been observed in a variety of natural systems, such as RNA hairpins\cite{Manosas:2006p1126, Liphardt:2001p2264} and ubiquitin\cite{Chyan:2004p2274} as well as specially designed molecules such as 
calixarene catenanes\cite{Janke:2008subm}.
In these systems fluctuations are observed due to the statistical nature of the transitions between a 'closed-bond' and an 'open-bond' conformation.
While in the context of SMF it is standard to analyze the fluctuations in the photon counts a treatment of the mentioned fluctuations in DFS is missing.

In the present letter we develop the theory for the analysis of the fluctuations observed in  FCS experiments of systems that show reversible bond-breaking events. 
We consider unbinding/rebinding transitions between two states, a 'closed-bond' state $A$ and an 'open-bond' state $B$ with rates $\koff(\fext)$ for $A\to B$-transitions and 
$\kon(\fext)$ for the reverse transition, where $\fext$ denotes the externally applied force.
While for time-independent rates $\koff(\fext)$ and $\kon(\fext)$, the process is Markovian, we additionally consider non-Markovian fluctuations.
SMF has proven to be a versatile tool to observe deviations from Markovian behavior, for a review see \cite{Barkai:2004p3178}.
In the context of DFS a brief discussion of non-Markovian fluctuations has been given by Hyeon and Thirumalai\cite{Hyeon:2007p2544}.
The impact of such fluctuations on experimental results apparently has not been considered in a quantitative manner up to now.
However, non-Markovian fluctuations can be expected to be of importance in particular when considering the dynamics of protein-unfolding and refolding due to the complex nature of this phenomenon\cite{Frauenfelder:1991p3198}.

In order to be able to describe the statistics of the unbinding/rebinding events, we consider the probability of finding the system in either state together with the number of transitions that have already occurred at time $t$, cf. Fig.1.

For the formal treatment, one considers the process consisting of the two variables 
$\{q(t),m(t)\}$, where $q(t)$ denotes the reaction coordinate and $m(t)$ the number of transitions. 
In the two-state approximation the distribution is given by 
$p(q,m;t)\simeq n_A(m;t)p_A^{\rm eq}(q)+n_B(m;t)p_B^{\rm eq}(q)$ with the equilibrium populations $p_X^{\rm eq}(q)$, $X=A$, $B$.
Here, we assumed that the intra-well relaxation is fast compared to the time scale of $\koff$ and $\kon$ (adiabatic approximation).
The quantities $n_X(m;t)$ obey the master equation
\cite{Brown:2003p1844,Brown:2006p1850}:
\Be\label{ME.n.xm}
\partial_t{n_A(m;t)}&=&-\koff(\fext) n_A(m;t)+\kon(\fext) n_B(m-1;t)\nonumber\\
\partial_t{n_B(m;t)}&=&-\kon(\fext) n_B(m;t)+\koff(\fext) n_A(m;t)
\Ee
In this form, the fluctuations in the number of transitions become evident. 
In particular, the marginal distributions 
$n_X(t)=\sum_{m=0}^\infty n_X(m;t)$ and $\bar p(m;t)=\int\!dqp(q,m;t)$ can be computed. 
The global populations $n_X(t)$ are those considered in the mentioned previous 
treatments\cite{Seifert:2002p163,Li:2006p164,Diezemann:2008subm} and $n_A(t)$ is related to the survival probability 
$p(\t)=-dn_A(\t)/d\t=\koff(\fext)\exp{(-[\koff(\fext)+\kon(\fext)]\t)}$.
It is to be noted that $g(\t)=\t p(\t)$ shows strong deviations from Poisson behavior if 
$\kon(\fext)\sim\koff(\fext)$, i.e. if the equilibrium constant 
\be\label{Keq.def}
\Keq(\fext)={\kon(\fext)\over\koff(\fext)}
\ee
is near unity. In addition, the correlation time, 
$\lg\t\rg=\int_0^\infty\!dt[p(t)t]
=\koff(\fext)/\left[\koff(\fext)+\kon(\fext)\right]^2$, shows a maximum as a function of the force.
In the Bell-limit (large $f_A$ and $f_B$, cf. Fig.1c), 
$\koff^{\rm (Bell)}(f)=\koff^0e^{\a_A f}$ and
$\kon^{\rm (Bell)}(f)=\kon^0e^{-\a_B f}$, this maximum is located at
$f_\t=\ln{[\Keq(0)(1+2\a_B/\a_A)]}/(\a_A+\a_B)$.
For the parameters given in the caption to Fig.1, one finds in this approximation 
$f_\t\simeq 45.3$ pN, while for the general Kramers rates
$\koff(f)=\koff^0e^{\a_A f(1-f/f_A)}$ and $\kon(f)=\kon^0e^{-\a_B f(1+f/f_B)}$ one has a value of $41.4$ pN, a deviation of 10\%.

The subsequent analysis of eq.(\ref{ME.n.xm}) proceeds in the same way as in 
ref.\cite{Brown:2006p1850}. â
We introduce generating functions
\be\label{Gx.def}
G_X(z;t)=\sum_{m=0}^\infty z^m n_X(m;t)
\quad(X=A,\, B)
\ee
from which all quantities of interest can be calculated\cite{vanKampen:1981}.
In the present letter we focus on the first and second moments of the distribution of the number of $A\leftrightarrow B$-transitions, $\lg N(t)\rg$ and $\lg N(t)^2\rg$. 
The latter will be discussed in terms of Mandels Q-parameter\cite{Barkai:2004p3178},
$Q(t)=\left[\lg N(t)^2\rg-\lg N(t)\rg^2\right]/\lg N(t)\rg-1$. 
This quantity will be shown to be highly susceptible to subtle changes ion the energy landscape. 
As described in detail in ref.\cite{Brown:2006p1850}, the generating functions can be used to calculate these quantities.
Starting in state $A$ initially, $G(z;t)=G_A(z;t)+G_B(z;t)$ reads:
\be\label{GF.result}
G(z;t)=\left[{W(z)+\Keq+1\over 2W(z)}E_1(z;t)+{W(z)-\Keq-1\over 2W(z)}E_2(z;t)\right]
\ee
with
$W(z)=\sqrt{(\Keq-1)^2+4z\Keq}$, 
$E_{1/2}(z;t)=
\left\lg\exp{\left(-\rho_{1/2}(z)\int_0^t\!dt'\koff(\fext,t')\right)}\right\rg$
and $\rho_{1/2}(z)={1\over2}\left[(\Keq+1)\mp W(z)\right]$. 
Here, we already allowed for a possible time-dependence of $\koff(\fext,t)$ due to environmental fluctuations and the brackets indicate the average over the disorder.
However, in this case, the employed formalism can only be used if we assume a unique 
$\Keq(\fext)=\kon(\fext,t)/\koff(\fext,t)$ to exist. 
For Markovian fluctuations, one simply has
$E_{1/2}(z;t)=\exp{\left(-\rho_{1/2}(z)\koff(\fext)t\right)}$.

The mean number of transitions $\lg N(t)\rg$ after a short time (on the order of the inverse equilibrium constant), where it depends on the actual initial conditions, grows linearly with slope
\be\label{R.def}
R=\lim_{t\to\infty}{d\over dt}\lg N(t)\rg
={\koff(\fext)\kon(\fext)\over \koff(\fext)+\kon(\fext)}
\ee
For the parameters given in the caption to Fig.1, we plot $\lg N(t)\rg$ versus time in the upper panel of Fig.2 along with results from a kinetic Monte Carlo simulation of the same system\cite{Rief:1998p1113}. 
It is evident that also for times on the order of $\koff$ one can observe the linear behavior. The slope $R$ according to eq.(\ref{R.def}) is given as a function of the external force in the inset. 
It shows a maximum as a function of $\fext$, which in the Bell-limit is located at $f_R=\ln{[\Keq(0)(\a_A/\a_B)]}/(\a_A+\a_B)$ (deviations from Kramers results $\sim$ 10\%).

The $Q$-parameter tends towards a constant value for long times:
\be\label{Q.limit}
Q_\infty=\lim_{t\to\infty}Q(t)
=-2{\Keq(\fext)\over \left[\Keq(\fext)+1\right]^2}
\ee
as shown in the lower panel of Fig.2.
Again, the Monte Carlo simulation results coincide with the analytical calculation, showing that $Q(t)$ can be extracted accurately from noisy (experimental) data. 
The minimum in $Q_\infty$ in the Bell-limit is located at $f_Q=\ln{[\Keq(0)]}/(\a_A+\a_B)$ 
(deviations from Kramers results $\sim$ 20\%).

These considerations show that it should be possible to extract both, $\lg N(t)\rg$ and $Q(t)$ from experimental data. 
For carrying out force-clamp experiments it is, however, highly advisable to carefully adjust filters that are meant to reduce instrumental noise but might suppress fluctuations that essentially contain valuable information about the inherent energy landscape.
The advantage of FCS is given by the fact that in addition to temperature the external force can be varied in order to adjust $\Keq(\fext)$ to a prescribed value.
Together with the information that can be gathered from the analysis of rupture force distributions\cite{Diezemann:2008subm}, one obtains a very detailed view about the structure of the energy landscape underlying the observed unbinding/rebinding transition.

We now show that FCS also is a versatile tool to extract information about the energy landscape in the situation where the fluctuations cannot be considered as Markovian. 
Hence, FCS can also be used to monitor the impact of environmental fluctuations or dynamic heterogeneities in the reaction under consideration. 
To this end we consider two models for dynamic disorder, namely a two-configuration exchange model (TCM) and a model of gaussian environmental fluctuations (GM)\cite{Brown:2006p1850}. 
In the TCM it is assumed that for each state $A$ and $B$ there are two configurations 
$(A,a)$, $(A,b)$ and $(B,a)$, $(B,b)$. 
The exchange between configurations $(X,a)$ and $(X,b)$ takes place with a rate $\G$. 
In a realistic treatment one would assume that $\G$ has a force-dependence that is different from that of the reaction rates. 
For simplicity, we neglect the force-dependency of $\G$ completely.
The correxponding reaction rates are $\koff^\a(\fext)$ for $(A,\a)\to(B,\a)$ and 
$\kon^\a(\fext)$ for the rebinding transition ($\a=a$, $b$). 
We assume that $\koff^\a(\fext)=x_\a\koff(\fext)$ and 
$\kon^\a(\fext)=x_\a\kon(\fext)$ where $\koff(\fext)$ and $\kon(\fext)$ are the rates with the parameters given in Fig.1 and $x_\a$ are constants.
In the GM, one assumes a continuous environmental variable $r(t)$ that undergoes a Ornstein-Uhlenbeck process with (force-independent) damping rate $\g$ and $\lg r^2\rg=\s^2$.
One furthermore chooses $\koff(\fext,r)=(r^2/\s^2)\koff(\fext)$, which allows an analytical treatment, cf. ref.\cite{Zwanzig:1990p2434}. 
In the context of SMF such gaussian models of dynamical disorder have been studied also including non-Markovian fluctuations\cite{Wang:1995p3155}. 
The calculations of $\lg N(t)\rg$ and $Q(t)$ can be performed 
analytically\cite{Brown:2006p1850} with the result that $\lg N(t)\rg$ is not affected at all by environmental fluctuations in the long-time limit.
For $Q(t)$ the situation is completely different.
For finite $\G$ and $\g$, one finds
\be\label{Q.TCM.GM}
Q_\infty=Q_\infty^{\rm (Markov)}+{\cal X}{\Keq(\fext)\over\Keq(\fext)+1}
\ee
where $Q_\infty^{\rm (Markov)}$ is given by eq.(\ref{Q.limit}) and one has
\be\label{X.def}
{\cal X}_{\rm TCM}=\left({1\over 2}{(x_a-x_b)^2\over x_a+x_b}\koff(\fext)\right)\G^{-1}
\quad;\quad
{\cal X}_{\rm GM}=\left(2\koff(\fext)\right)\g^{-1}
\ee
The important feature of eq.(\ref{Q.TCM.GM}) is that for finite $\G$ ($\g$) $Q_\infty$ becomes positive! 
This is a clearcut indication of non-Markovian fluctuations since $Q_\infty$ is always negative (or zero) for Markovian fluctuations.
It is evident, that only for $\G\to\infty$ ($\g\to\infty$), i.e. in the motional narrowing regime, the Markovian limit is recovered. 
In the limit of static disorder, $\G\to 0$ ($\g\to 0$), $Q(t)$ does not reach a limiting value in a finite time.
Instead, $Q(t)$ increases linearly with time and can be written as $Q(t)=\hat{\cal X}t$. 
The rate $\hat{\cal X}$ is obtained from eq.(\ref{X.def}) by replacing $\G$ ($\g$) with unity. 
Moreover, eq.(\ref{X.def}) shows that $Q(t)$ does not allow to discriminate between different scenarios. For the two models considered only the time scale on which the limit is reached differs somewhat.

In Fig.3 we plot $Q_\infty$ as a function of the external force for the GM.
In order to show that this result also is relevant for the interpretation of experimental data, we show in Fig.4 how the positive limit is reached as a function of time.
Again, for not too small $\g$, $Q_\infty$ is reached on the time scale of $\koff$.

We have scrutinized the results of typical FCS experiments performed on systems showing reversible bond breakage. 
The fluctuations in such systems can be analyzed in much the same way as the photon counts in SMF. However, force spectroscopy allows to manipulate the reaction rates and therefore one can obtain information about the energy landscape that is not available from other single molecule techniques. 
Subtle features of the conformational dynamics displayed by reversibly bonded systems are revealed which allows to expand the versatility of force spectroscopy in terms of fluctuation analysis.
We have also shown that it should be possible to discriminate between Markovian and non-Markovian fluctuations by examining Mandel's $Q$-parameter - a parameter which is experimentally accessible. 
We expect that experiments analyzed along the way discussed in this work will yield valuable new information regarding the details of complex chemical reactions under external force.
\section*{Acknowledgment}
This work has been supported by the Deutsche Forschungsgemeinschaft via SFB 625.
%

%
\section*{Figure captions}
\begin{description}
%
\item[Fig.1]
{\bf a)}: Sketch of the envisioned transitions between the bond (A) and the unbound state (B);\\
{\bf b)}: A typical force-clamp trajectory from a simulation of Brownian dynamics in a double-well potential\\
{\bf c)}: Reaction rates in Kramers-approximation as function of external force;
explictly, one has $\koff(f)=\koff^0e^{\a_A f(1-f/f_A)}$ and 
$\kon(f)=\kon^0e^{-\a_B f(1+f/f_B)}$.
We choose for the parameters: $\a_A/\b=0.3nm$, $\a_B/\b=0.7nm$, $f_A=800pN$, $f_B=200pN$. The rates for zero force are chosen as $\koff^0=1s^{-1}$ and $\kon^0=10^4s^{-1}$. 
The dotted lines are the Bell-limit for the rates, i.e. the limit of large $f_A$ and 
$f_B$ of the above expressions: $\koff^{\rm (Bell)}(f)=\koff^0e^{\a_A f}$ and
$\kon^{\rm (Bell)}(f)=\kon^0e^{-\a_B f}$.
$\koff^0$ sets the time scale for all calculations.
\item[Fig.2] (color online)\\
upper panel: Mean number of transitions versus time for the same parameters as in Fig.1. The dots are results of kinetic Monte Carlo simulations.
Inset: Slope $R$ versus external force.\\
Lower panel: $Q(t)$ versus time for the same parameters as in Fig.1. 
Inset: $Q_\infty$ versus external force.
\item[Fig.3] (color online)\\
$Q_\infty$ versus $\fext$ for the same parameters as in Fig.1.
The environmental fluctuations are modelled by the GM.
\item[Fig.4] (color online)\\
$Q(t)$ versus $\fext$ for the same parameters as in Fig.1.
The environmental fluctuations are modelled by the GM with $\g=10$s$^{-1}$.
The lines are for different $\Keq(\fext)$, from bottom to top:
$0.1$, $0.2$, $0.5$, $1$ (red), $2$, $5$ and $10$.
\end{description}
\newpage
\begin{figure}
\centering
\includegraphics[width=18cm]{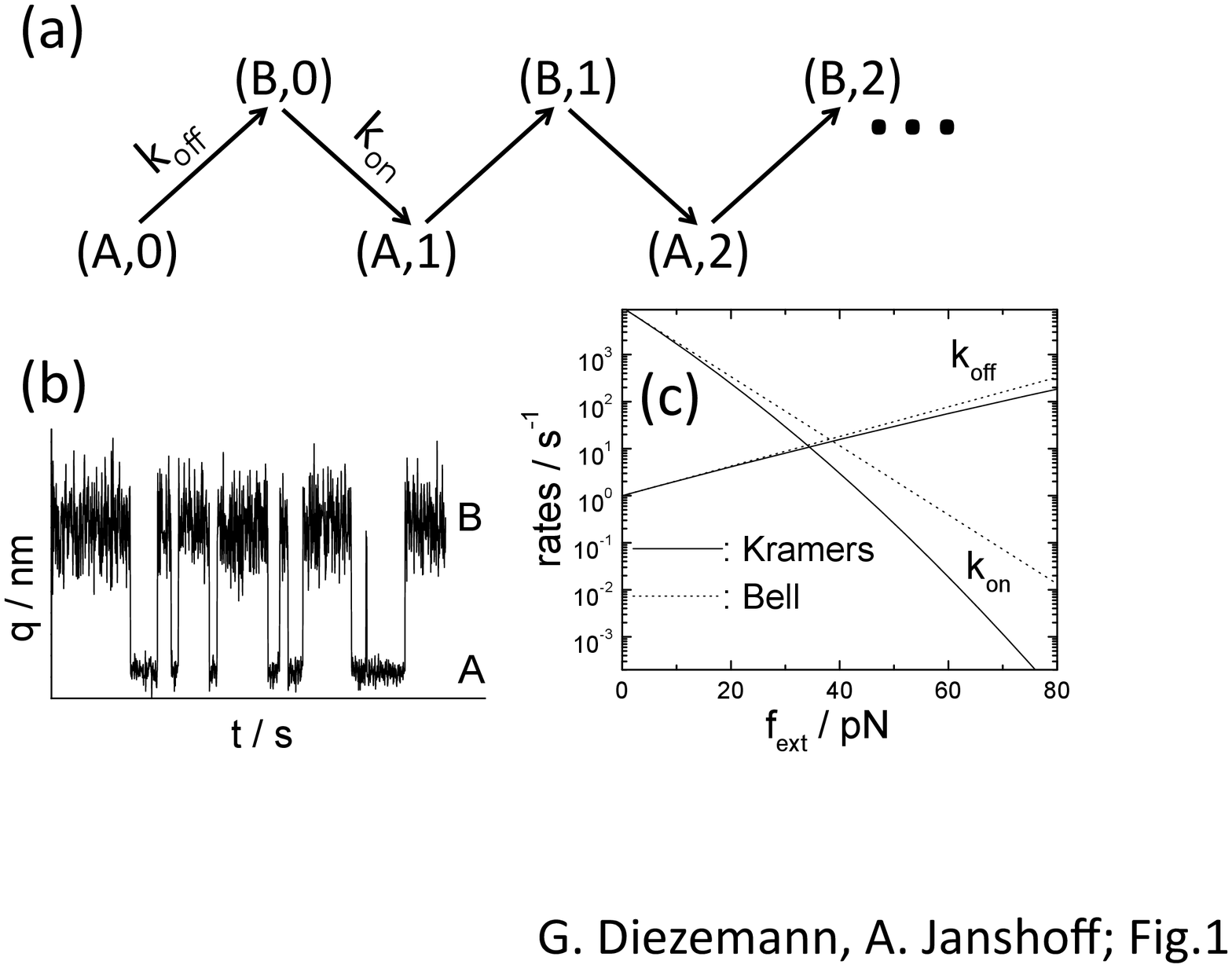}
\end{figure}
\newpage
\begin{figure}[h!]
\centering
\includegraphics[width=15cm]{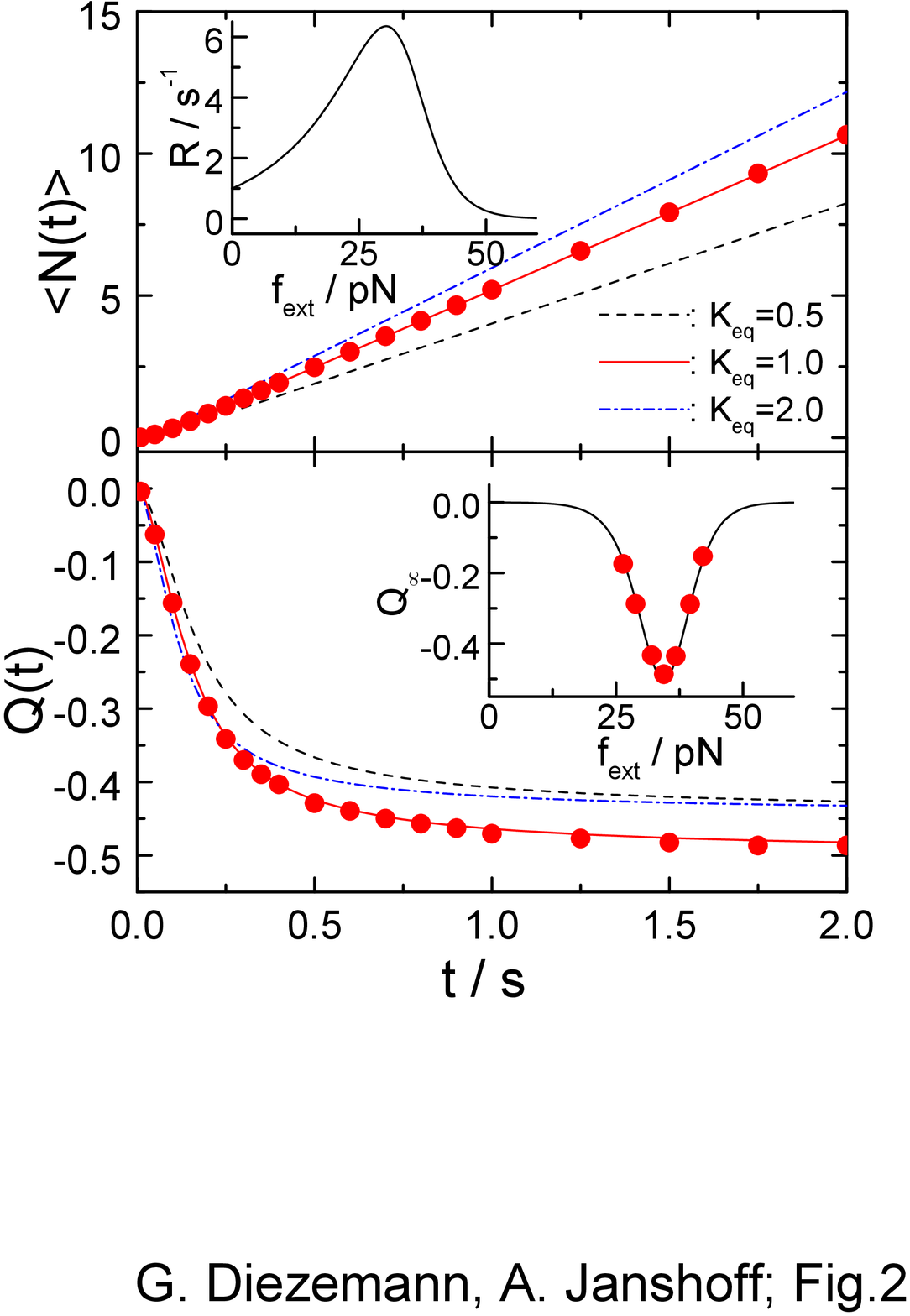}
\end{figure}
\newpage
\begin{figure}[h!]
\centering
\includegraphics[width=15cm]{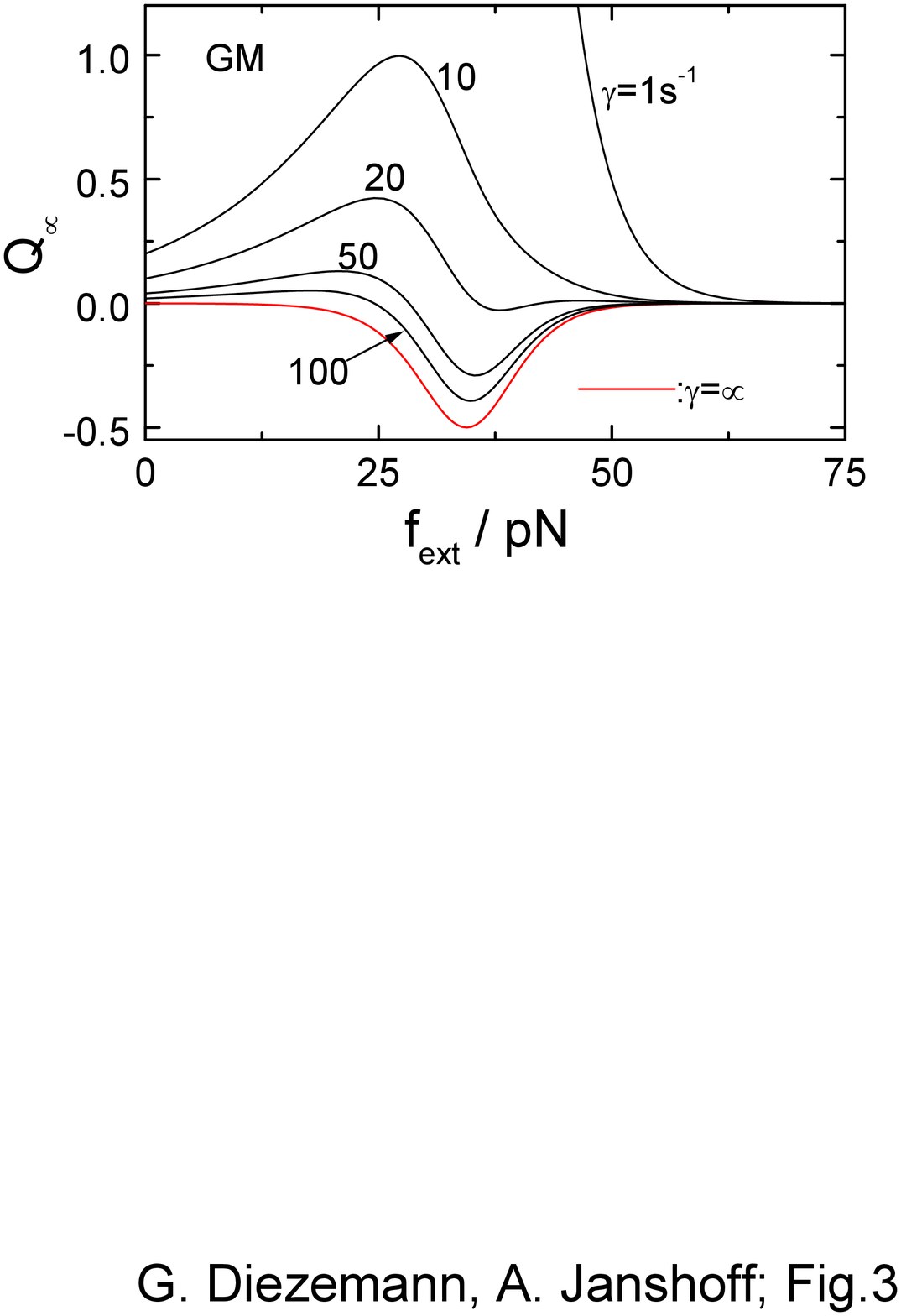}
\end{figure}
\newpage
\begin{figure}[h!]
\centering
\includegraphics[width=15cm]{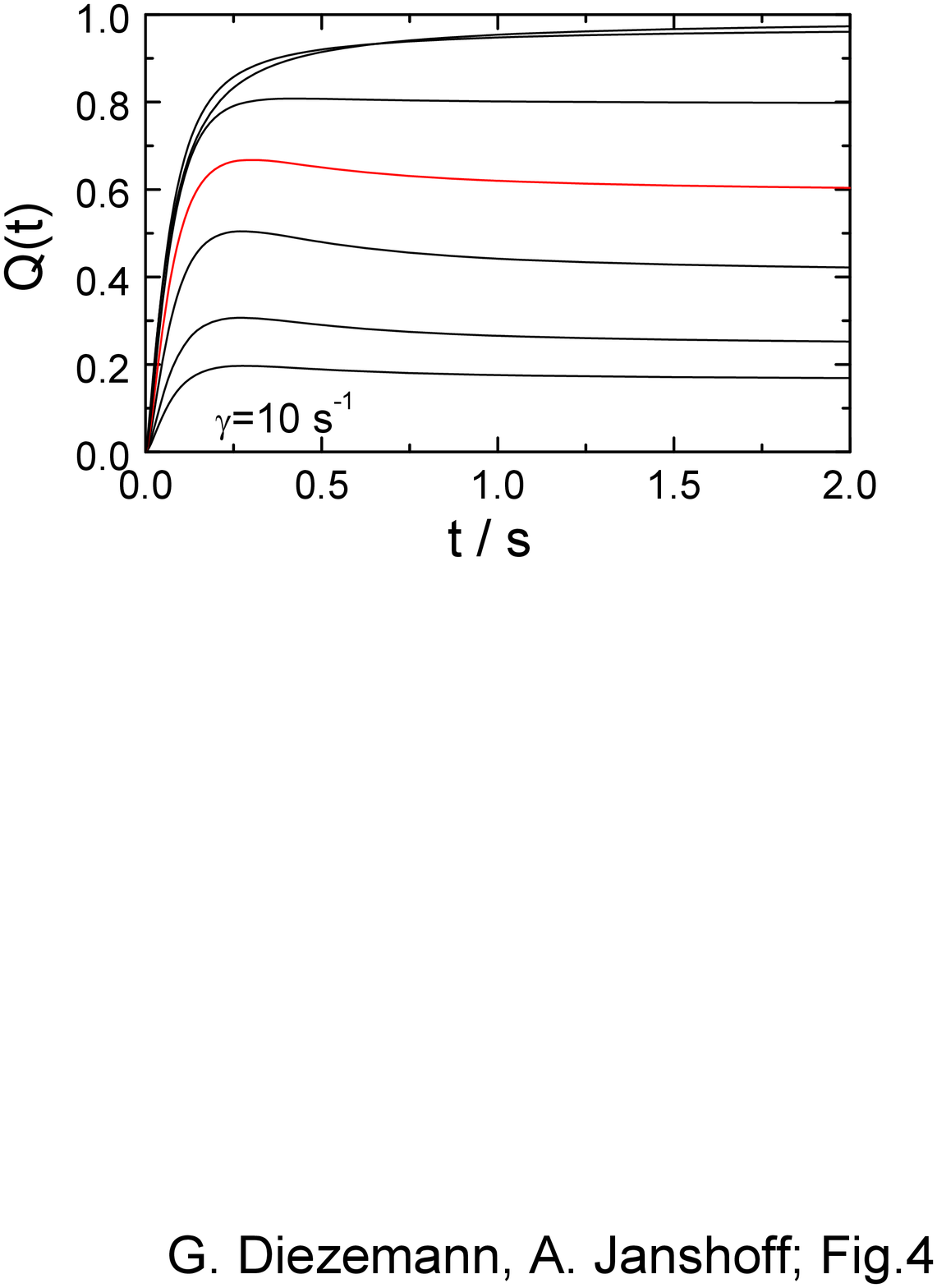}
\end{figure}

\end{document}